\documentclass[aps,prl,twocolumn,showpacs,10pt,superscriptaddress,preprintnumbers,nofootinbib,floatfix]{revtex4-1}

\usepackage{bm}
\usepackage{graphicx}
\usepackage{amsmath,amssymb,mathtools} % Math
\usepackage{xspace}
\usepackage{wrapfig}
\usepackage{subfig}
\usepackage[format=plain,justification=Justified]{caption}

\usepackage{slashed}
\usepackage[normalem]{ulem}
\usepackage[dvipsnames]{xcolor}
\usepackage[utf8]{inputenc}
\usepackage[T1]{fontenc}
\usepackage{mathtools}
\usepackage{stmaryrd}
\usepackage{appendix}
\usepackage{xspace}
\usepackage{upgreek}
\usepackage{multirow}
\usepackage[textsize=tiny]{todonotes}
\usepackage{enumitem}

\usepackage{scrextend}
\usepackage[colorlinks=true
,urlcolor=RoyalBlue
,anchorcolor=RoyalBlue
,citecolor=RoyalBlue
,filecolor=RoyalBlue
,linkcolor=RoyalBlue
,menucolor=RoyalBlue
,linktocpage=true
,pdfa=true]{hyperref} 

\usepackage{tabularx}
\usepackage{tensor}

\newcommand{\mb}[1]{\boldsymbol{#1}}

\usepackage{marginnote}

\definecolor{darkgreen}{rgb}{0.0, 0.4, 0.0}

\newcommand{\as}{\alpha_s}

\newcommand{\nn}{\nonumber}

\newcommand{\Pythiaxx}{\texttt{Pythia\xspace8.3}\xspace}
\newcommand{\Pythia}{\texttt{Pythia}\xspace}
\newcommand{\Vincia}{\texttt{Vincia}\xspace}
\newcommand{\Vinciaxx}{\texttt{Vincia\xspace2.3}\xspace}
\newcommand{\Herwig}{\texttt{Herwig}\xspace}
\newcommand{\Herwigxx}{\texttt{Herwig\xspace7.3}\xspace}

\newcommand{\Fastjet}{\texttt{FastJet}\xspace}

\newcommand{\letter}{\emph{Letter}\xspace}

\def\nn{\nonumber}

\allowdisplaybreaks[2]

\newcommand{\cO}{\mathcal{O}}

\DeclareRobustCommand{\Refcite}[1]{Ref.~\cite{#1}}

\DeclareRobustCommand{\Refscite}[1]{Refs.~\cite{#1}}

\DeclareRobustCommand{\eq}[1]{eq.~\eqref{eq:#1}}

\DeclareRobustCommand{\eqs}[2]{eqs.~\eqref{eq:#1} and \eqref{eq:#2}}

\DeclareRobustCommand{\secn}[1]{\hyperref[sec:#1]{section~\ref*{sec:#1}}}
\DeclareRobustCommand{\Sec}[1]{\hyperref[sec:#1]{Section~\ref*{sec:#1}}}

\DeclareRobustCommand{\subsec}[1]{\hyperref[subsec:#1]{subsection~\ref*{subsec:#1}}}
\DeclareRobustCommand{\Subsec}[1]{\hyperref[subsec:#1]{Subsection~\ref*{subsec:#1}}}

\DeclareRobustCommand{\app}[1]{\hyperref[app:#1]{appendix~\ref*{app:#1}}}
\DeclareRobustCommand{\App}[1]{\hyperref[app:#1]{Appendix~\ref*{app:#1}}}

\DeclareRobustCommand{\fig}[1]{\hyperref[fig:#1]{Figure~\ref*{fig:#1}}}
\DeclareRobustCommand{\Fig}[1]{\hyperref[fig:#1]{Figure~\ref*{fig:#1}}}
\DeclareRobustCommand{\figs}[2]{Figs.~\ref{fig:#1} and \ref{fig:#2}}

\DeclareRobustCommand{\tab}[1]{\hyperref[tab:#1]{table~\ref*{tab:#1}}}
\DeclareRobustCommand{\Tab}[1]{\hyperref[tab:#1]{Table~\ref*{tab:#1}}}

\newcommand{\adi}[1]{\textcolor{RoyalBlue}{#1}}

\newcommand{\td}{\mathrm{d}}

\begin{document}

\title{Using the $\boldsymbol{W}$ as a Standard Candle to Reach the Top:\\ Calibrating Energy Correlator Based Top Mass Measurements}

\author{Jack Holguin}
\affiliation{University of Manchester, School of Physics and Astronomy, Manchester, M13 9PL, United Kingdom}
\affiliation{CPHT, CNRS, \'Ecole polytechnique, Institut Polytechnique de Paris, 91120 Palaiseau, France}

\author{Ian Moult}
\affiliation{Department of Physics, Yale University, New Haven, CT 06511}

\author{Aditya Pathak}
\affiliation{Deutsches Elektronen-Synchrotron DESY, Notkestr. 85, 22607 Hamburg, Germany}

\author{Massimiliano Procura}
\affiliation{University of Vienna, Faculty of Physics, Boltzmanngasse 5, A-1090 Vienna, Austria}

\author{Robert Sch\"ofbeck}
\affiliation{Institute for High Energy Physics, Austrian Academy of Sciences, Nikolsdorfergasse 18, A-1050 Vienna, Austria}

\author{Dennis Schwarz}
\affiliation{Institute for High Energy Physics, Austrian Academy of Sciences, Nikolsdorfergasse 18, A-1050 Vienna, Austria}

\preprint{UWThPh 2023-26}
\preprint{DESY-23-176}

\begin{abstract}

The top quark mass is a key parameter of the Standard Model, yet measuring it precisely at the Large Hadron Collider (LHC) is challenging. Inspired by the use of standard candles in cosmology, we propose a novel energy correlator-based observable, which directly accesses the dimensionless quantity $m_t$/$m_W$. We perform a Monte Carlo study to demonstrate the feasibility of the top mass extraction from Run 2, 3, and High-Luminosity LHC datasets. Our resulting $m_t$ can be defined in a well-controlled short-distance mass scheme and exhibits remarkably small uncertainties from nonperturbative effects, as well as insensitivity to parton distribution functions, outlining a roadmap for a record precision measurement at the LHC.

\end{abstract}

\maketitle

\emph{Introduction.}---The precise value of the top quark mass plays a crucial role in the Standard Model, both for testing its internal consistency \cite{Baak:2012kk,Baak:2014ora}, and for understanding its extrapolation to high energies for hints of what lies beyond \cite{Espinosa:2007qp,Arkani-Hamed:2008mpk,Elias-Miro:2011sqh,Degrassi:2012ry,Buttazzo:2013uya,Bezrukov:2014ina,Espinosa:2015qea,Giudice:2015toa,Steingasser:2023ugv,Khoury:2021zao}. However, the top can currently only be produced at hadron colliders, where precision measurements of its mass are notoriously difficult. While a large number of techniques have been proposed and used in precision measurements \cite{CDF:2014upy,CMS:2015lbj,ATLAS:2016muw,ParticleDataGroup:2020ssz}, they all suffer from difficulties, such as the lack of theoretical control over the observables' relation to the Lagrangian top mass parameter~\cite{Nason:2017cxd,Hoang:2020iah}, or a high sensitivity to parton distribution functions (PDFs). With the High-Luminosity Large Hadron Collider (HL-LHC) era on the horizon, a renewed effort to develop qualitatively new approaches to extract the top mass is warranted.

A novel approach to studying complex final states in hadronic decays is jet substructure (e.g.~\Refscite{Larkoski:2017jix,Kogler:2018hem,Marzani:2019hun} and references therein). Jet substructure enables the study of decaying heavy particles through their imprint in the correlations of hadronic energy flux within jets.
As first shown in \Refcite{Hoang:2017kmk}, this provides an opportunity to study inclusive decays of boosted top quarks at the LHC, which can be described with rigorous factorization theorems~\cite{Fleming:2007xt,Fleming:2007qr}, enabling the top mass to be treated in short-distance mass schemes~\cite{Hoang:2009yr,Hoang:2017suc}, overcoming ambiguities in its definition present in other approaches.

So far, two observables have been put forward for precision extractions of the top mass from jet substructure: one dimensionful and one dimensionless. The
first approach~\cite{Hoang:2017kmk}, further developed in \Refscite{Hoang:2019ceu,Bachu:2020nqn},
proposes to use the groomed~\cite{Dasgupta:2013ihk,Larkoski:2014wba} jet mass. This is directly sensitive to the dimensionful mass scale of the top. However, it is also sensitive to soft radiation and nonperturbative effects at scales parametrically lower than the top width, $\Gamma_t$. This can be ameliorated by grooming but at the
expense of an increased theoretical complexity.
While progress has been made in understanding the description of nonperturbative corrections for groomed jets~\cite{Hoang:2019ceu,Ferdinand:2023vaf,Pathak:2023sgi}, the distribution also receives a large contribution from the underlying event (UE) for the grooming proposed in \Refcite{Hoang:2017kmk}, which is extremely challenging to model accurately.
The second proposal~\cite{Holguin:2022epo} is to use energy correlators (EECs)~\cite{Basham:1978bw,Basham:1977iq,Basham:1979gh,Basham:1978zq,Hofman:2008ar}, which have recently
been introduced~\cite{Chen:2020vvp} as jet substructure observables.
EECs were first measured in hadron colliders in \Refscite{CMS:2024mlf,Tamis:2023guc,talk1}.
Unlike the jet mass, EECs are sensitive to a dimensionless \emph{angular scale}, with the mass scale being reconstructed from the jet $p_T$. Since soft radiation enters the energy correlators only via recoil, they are not sensitive to scales below $\Gamma_t$, making them naturally robust to both hadronization and UE.
Despite the theoretical elegance of this approach, the jet $p_T$ has large experimental uncertainties, making a precise determination of $m_t$ challenging in practice. We thus believe that identifying a top-mass-sensitive observable that is simultaneously experimentally
feasible at the LHC, robust to hadronization and UE, and calculable to high perturbative orders
remains an important open problem.

In this \emph{Letter}, we introduce an EEC-based observable for precision top quark mass measurements, which overcomes the previous experimental difficulties. Our observable is inspired by cosmology, where it is common that precisely measured observables, such as luminosity, are not directly related to quantities of interest, such as distances. The use of standard candles then plays a crucial role, providing a methodology for converting between two independent dimensionful quantities. This is similar to the present case of extracting masses from measurements of high-multiplicity hadronic states: the dimensionless angular scales are robust observables, necessitating the development of standard candles to enable their use for precision mass measurements. Crucially, the top predominately decays into an electroweak scale particle whose mass has been measured with spectacular accuracy, the $W$ boson. This particle provides the needed standard candle by introducing another dimensionless parameter, $m_t/m_W$, into the observable. In this \emph{Letter}, we study a hadronization and UE insensitive standard candle constructed from EECs measured on the $W$ boson, allowing us to build a distance ladder all the way back through the complicated QCD dynamics to the time scales of the top quark decay. The outcome is a measurement of the top mass in terms of the $W$ mass and a purely angular measurement which are both independently robust experimental quantities.
We emphasize that this approach is distinct from existing top mass extractions \cite{CMS:2022kqg,CMS:2023ebf} wherein the $W$ decay is reconstructed only to achieve a fine-grained calibration of the jet energy scale (JES).
These methods are otherwise not a priori robust to experimental and modelling uncertainties. In contrast, our approach offers a higher potential for precision since we work with angular quantities that are already experimentally robust from the outset and are more amenable to theoretical calculations. This is supported by recent studies~\cite{Holguin:2024tkz,Xiao:2024rol}.

We here demonstrate the feasibility and properties of our approach at the LHC through a Monte Carlo study and lay out a roadmap for an experimental and theoretical program to achieve a record top mass measurement.

\begin{figure}
\centering
\subfloat[The shape of the three-point correlator on boosted top jets, \eq{observable}. A large value of $\zeta_S$ selects the hard top decay process, but by lowering $\zeta_S$, the $W$ peak emerges. Slices for specific values of $\zeta_S$ are shown on the boundaries of the plot.]{\includegraphics[width=0.47\textwidth]{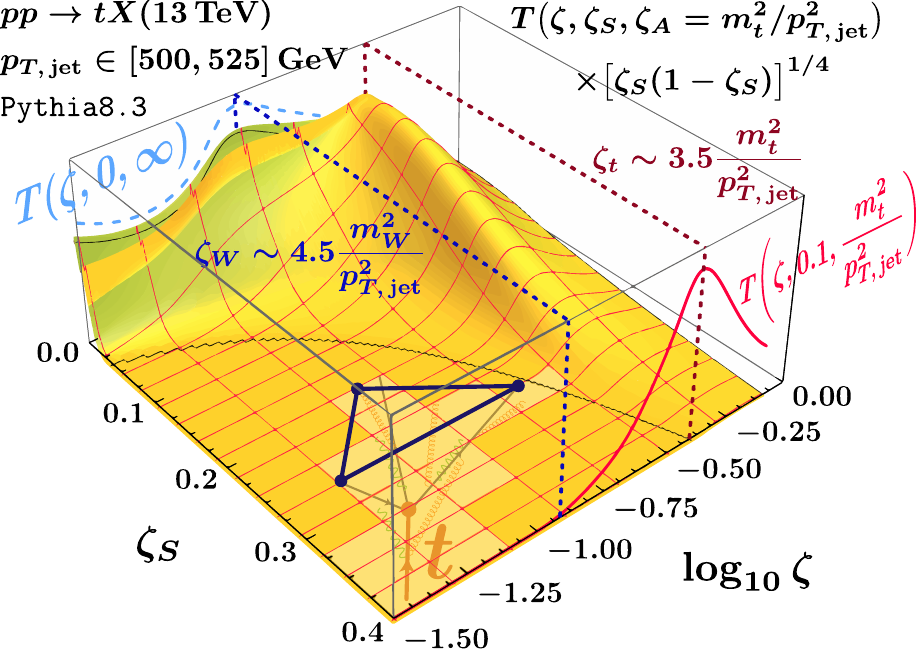}
\label{fig:armadillo_a}}\\
\subfloat[Slices for specific values of $\zeta_S$ which emphasize the sharpness of the $W$ and top peaks. The green line with the small bump corresponds to the equilateral projection considered in \cite{Holguin:2022epo}.]{\includegraphics[width=0.47\textwidth]{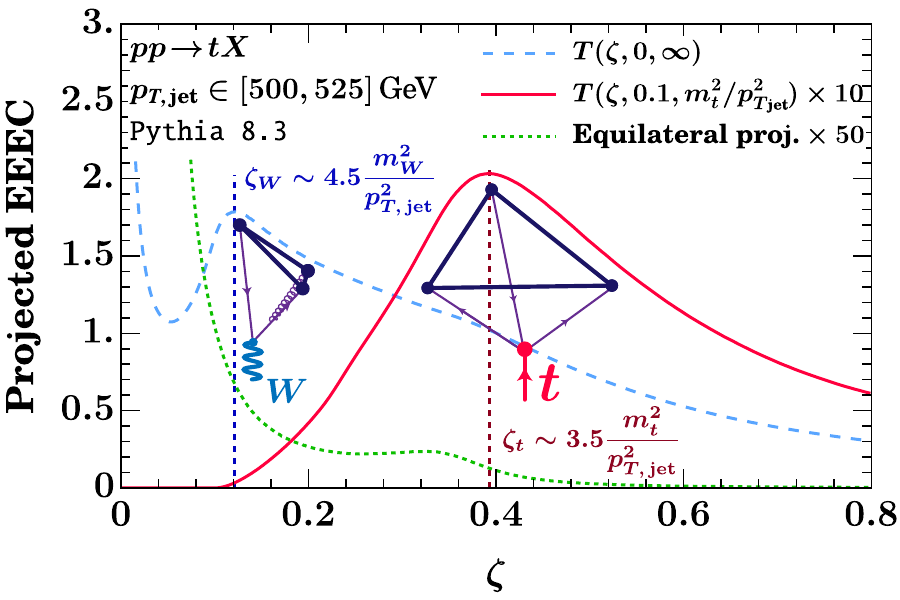}
\label{fig:armadillo_b}}
\caption{Illustrative plots produced from \Pythia showing the imprint of top and $W$ on the 3-point EEC in \eq{observable}.}
\label{fig:armadillo}
\end{figure}

\emph{Energy Correlators on Top Decays.}---
There has been rapid progress in our understanding of multi-point energy correlators and their application to jet substructure (e.g. \Refscite{Komiske:2022enw,Liu:2022wop,Liu:2023aqb,Cao:2023oef,Chen:2019bpb,Chen:2020adz,Chen:2022swd,Holguin:2022epo,Lee:2022ige,Craft:2022kdo,Devereaux:2023vjz,Jaarsma:2023ell,Lee:2023npz,Yang:2023dwc,Barata:2023zqg,Andres:2022ovj,Andres:2023xwr}), as well as their structure for generic angles~\cite{Yan:2022cye,Yang:2022tgm} and theoretical properties~\cite{Chen:2022jhb,Chang:2022ryc}. In \Refcite{Holguin:2022epo}, the three-point correlator was applied to detect the angular scale associated with the top decay. Since, at leading order, this is a hard three-body decay, it was proposed that this could be detected in an equilateral configuration for the correlator. However, the full three-point correlator on top decays is a rich function of three angles whose shape has not yet been explored.

The key object of our analysis is the following integrated EEC (weighted cross-section).
We express the angles between the momenta of the correlated final state particles as $\zeta_{ij}=\Delta \eta_{ij}^2 + \Delta \phi_{ij}^2$ in terms of rapidity-azimuth coordinates.
The observable we define is
\begin{align}
& T(\zeta, \zeta_S, \zeta_{A}) \equiv \sum_{{}^{\mathrm{hadrons}}_{~\, i,j,k}}\int \td \zeta_{ijk} ~
\frac{p_{T,i} \, p_{T,j} \, p_{T,k}}{\big(p_{T,\mathrm{jet}}\big)^{3}} ~ \frac{\td^{3} \sigma_{i,j,k}}{\td
\zeta_{ijk}} \nonumber \\
& \times \Theta(\zeta_{ij}\geq\zeta_{jk}\geq\zeta_{ki}\geq \zeta_{S}) ~ \delta\left(\zeta -
\Big(\frac{\sqrt{\zeta_{ij}}+\sqrt{\zeta_{jk}}}{2}\Big)^2 \right) \nonumber \\
& \times \Theta\left(\zeta_{A} >(\sqrt{\zeta_{ij}}-\sqrt{\zeta_{jk}})^2 \right)\,. \label{eq:observable}
\end{align}
Here the sum is over all (not necessarily distinct) triplets of hadrons $(i,j,k)$ in the most energetic jets within each hemisphere of an identified top event. In the last equation, $p_{T,i}$ is the transverse momentum of particle $i$, which provides the energy weighting, and $\td \zeta_{ijk} \equiv \td \zeta_{ij} \td \zeta_{jk} \td \zeta_{ki}$.
We find it convenient to trade the three angles of the correlator for an overall scale $\zeta$, a cut on the smallest angle in the correlator, $\zeta_S$, and a cut on the asymmetry between the longest two sides, $\zeta_A$. This parametrization of the three-point correlator allows us to easily control the phase-space regions identified by the correlator. A plot of this distribution on boosted top
quarks is shown in \fig{armadillo_a}, revealing an interesting shape. For large values of $\zeta_S$ there is a single peak at the angular scale of the hard three-body decay of the top quark, $\zeta_t\sim m_t^2/ p_{T,\mathrm{jet}}^2$. As $\zeta_S$ is lowered, corresponding to the inclusion of more squeezed triangle configurations, a peak at the angular scale of the $W$, $\zeta_W\sim m_W^2/ p_{T,\mathrm{jet}}^2$ emerges in $T$ and is clearly visible in the back wall of \fig{armadillo_a}. In this work, $\zeta_S$ and $\zeta_A$ are fixed at the particular values $\zeta_S= 0.8 (172 [\mathrm{GeV}]/p_{T \, \mathrm{jet}})^2$ and $\zeta_A= (172 [\mathrm{GeV}]/p_{T \, \mathrm{jet}})^2$ to cleanly isolate the top peak feature.
The distributions with fixed $\zeta_S$ cuts are shown on the boundaries of the plot in \fig{armadillo_a} and separately in \fig{armadillo_b}, which highlights the sharpness of the peaks, as well as the kinematic configurations that give rise to them. \fig{armadillo_b} also highlights the substantial statistical gain of $T(\zeta,\zeta_{S},\zeta_{A})$ relative to the near equilateral projection previously used in the literature \cite{Holguin:2022epo}. The topologies depicted in \fig{armadillo_b} are analogous to the ``equilateral'' and ``squeezed'' configurations for cosmological correlators \cite{Arkani-Hamed:2015bza,Arkani-Hamed:2018kmz}.
This shows that we can simultaneously resolve the $W$ and top inside a single jet without performing any reclustering, which is a remarkable feature of the EECs that are inclusively-sensitive to all scales present in a given system. This property of the EECs was first applied in \cite{Andres:2023ymw}, where they were used to resolve both the $b$-quark mass scale and the length scale of the plasma in heavy-ion collisions.

\begin{figure}
\centering
\includegraphics[width=0.47\textwidth]{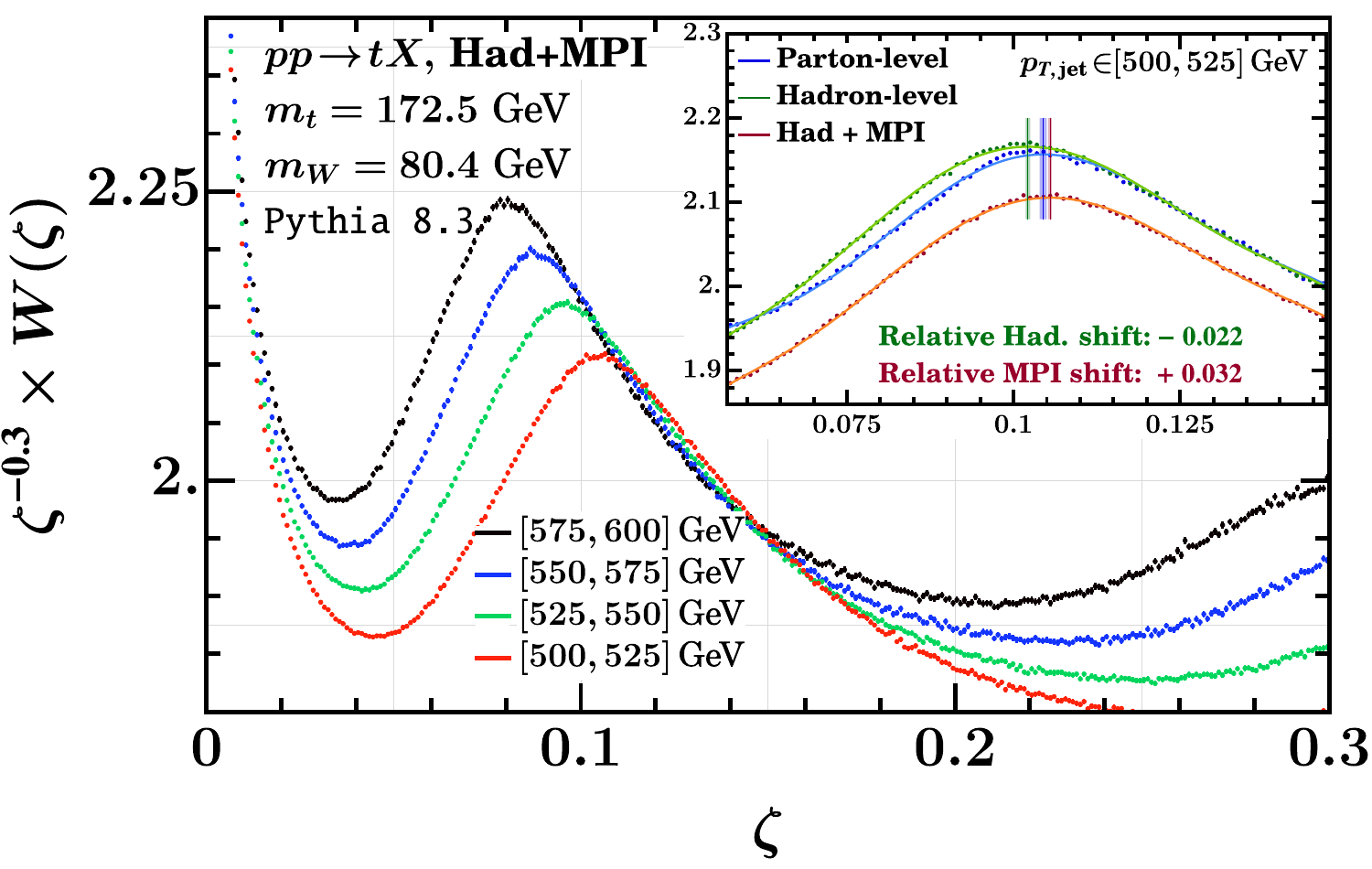}
\caption{The ratio of the three- to two-point correlator defined in Eq.~\eqref{eq:ratio} provides a robust standard candle identifying the angular scale of the $W$ mass. The peak is shown for different values of the jet $p_T$, and the inset shows the robustness of the shape of the peak region to hadronization and UE effects. The factor of $\zeta^{-0.3} \approx \zeta^{-\gamma(4)+\gamma(3)}$, involving the relevant spin 4 and 3 anomalous dimensions, is included to reduce the tilt from universal small-angle physics making the peak easier to fit.}
\label{fig:candle}
\end{figure}

\emph{The $W$ as a Standard Candle.}---The identification of the angular scale of the $W$ is the first key insight of this \emph{Letter}, and opens the door to using $m_W$, which is known with a remarkable uncertainty of $\pm 12$ MeV~\cite{ParticleDataGroup:2022pth}, to calibrate the mass scale of the top quark. The presence of the $W$ introduces a dependence of the EEC distribution on the dimensionless scale $m_t/m_W$, which can be extracted from measurements of the shape of the distribution around the peak. In particular, the {\emph{ratio}} of the location of its peaks is determined by $m_t/m_W$. This reflects the commonality between the rest frame of the top and the frame in which the $W$'s production is isotropic, and so their peak locations are determined by a single shared boost. The distribution in the peak region can be computed using rigorous factorization theorems and expressed in terms of the top mass in a short-distance mass scheme \cite{Hoang:2009yr,Hoang:2017suc}. The known value of $m_W$ then allows a conversion to a measurement of $m_t$ in a short-distance mass scheme.
In this light, the complicated three-body decay of the top, via the intermediate $W$, is in fact a gift.

To exploit the presence of the $W$ mass scale, we must develop a standard candle observable that is robust to hadronization and pile-up effects and which can be calculated to high perturbative orders. Unlike for the top, where the scale $\Gamma_t$ provides a natural cutoff suppressing nonperturbative contributions in the peak region, for the $W$ there is no such cutoff. We must suppress nonperturbative effects in another fashion. One approach to obtaining EEC observables that are robust to hadronization is to take ratios that cancel leading nonperturbative effects. This was originally proposed in \Refcite{Chen:2020vvp}, and these ratios have since been calculated to high perturbative orders~\cite{Lee:2022ige,Jaarsma:2023ell,Chen:2023zlx} and measured in collider experiments~\cite{Komiske:2022enw}. Here, we extend this approach to produce a robust standard candle from the $W$ boson.

We define the ratio
\begin{align}
\label{eq:ratio}
W(\zeta) &\equiv T(\zeta,0,\infty)\\
&~ \times\left(\sum_{{}^{\mathrm{hadrons}}_{~~ i,j}}\int \td \zeta_{ij} ~ \frac{p_{T,i} \,
p_{T,j}}{\big(p_{T,\mathrm{jet}}\big)^{2}} ~ \frac{\td \sigma_{i,j}}{\td \zeta_{ij}}~ \delta(\zeta -
\zeta_{ij} )\right)^{-1}, \nonumber
\end{align}
where the denominator is the standard two-point EEC, while the numerator, $T(\zeta,0,\infty)$, is the correlator of \eq{observable}, with no cuts on the asymmetry $\zeta_A$, or smallest angle, $\zeta_S$. Note that this observable is nearly identical to the standard ratio of projected correlators~\cite{Chen:2020vvp}, with the only difference being that we take $\zeta$ in the three-point EEC to be the average of the long and medium side, as compared to the longest side as done in \Refcite{Chen:2020vvp} since we find that this is better behaved in the peak region. When defined in this way, the two- and the three-point EEC in $W(\zeta)$ have closely related leading nonperturbative corrections in the peak region. Specifically, these nonperturbative effects partly arise from the (boosted) Sudakov region involving the same soft function in the numerator and denominator of the $W(\zeta)$, such that they cancel at leading logarithmic accuracy~\cite{Feige:2012vc}. Furthermore, the nonperturbative corrections of the kind that appear in the collinear limit of projected EECs cancel in the same way in the $W(\zeta)$ ratio as they do for massless jets~\cite{Chen:2020vvp}. \Fig{candle} shows this observable for different values of the jet $p_T$, illustrating the resulting peak structure. The inset shows shifts in the peak due to hadronization and the UE. We will show below that these shifts are highly correlated with those in the top quark peak, and arise solely due to shifts in the $p_{T, \mathrm{jet}}$ distribution.

In addition to its robustness, this observable can be computed to high perturbative orders in the peak region using a factorization theorem building on \Refscite{Moult:2018jzp,Gao:2019ojf}, which we leave to future work. Combined, these features provide us with a well-calibrated standard candle in the complex LHC hadronic environment. To our knowledge, such an approach has not previously been used in precision jet substructure measurements, and it would be interesting to explore if this could be applied in other situations.

\emph{Monte Carlo study.}---Having illustrated that EECs can be used to robustly identify the angular scales associated with the $W$ and top masses, we now show that this allows us to precisely extract the value of the top mass in units of the known $W$ mass. Crucially, we demonstrate that this calibration can be performed without needing to know the jet $p_T$, which has large experimental uncertainties, large nonperturbative corrections~\cite{Dasgupta:2007wa}, and is sensitive to the parton distribution functions (PDFs).

\begin{figure}
\centering
\subfloat[The ratio $\zeta_t/\zeta_W$ as a function of $p_{T,\,{\rm jet}}$ enabling the conversion between the standard candle and the top mass according to \eq{Cfactor}. While the ratio is virtually independent of the jet $p_T$, it is sensitive to $m_t$ and $m_W$.]{\includegraphics[width=0.47\textwidth]{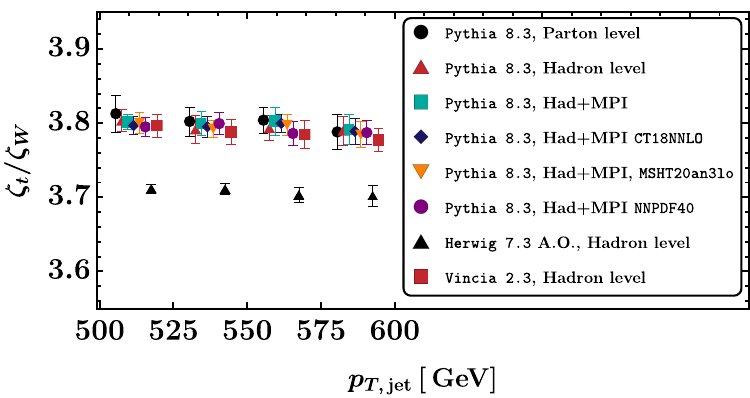}
\label{fig:conversion_a}}\\
\subfloat[The shift in the extracted top mass from simulation due to hadronization, MPI effects, and modification of the PDFs. Simulations are performed with \Pythia unless labeled otherwise. Variation of the PDF set is performed with hadronization and MPI. Averages across jet $p_{T}$ bins are shown in the legend. This illustrates the resilience of the standard candle to nonperturbative and initial-state effects.]{\includegraphics[width=0.48\textwidth]{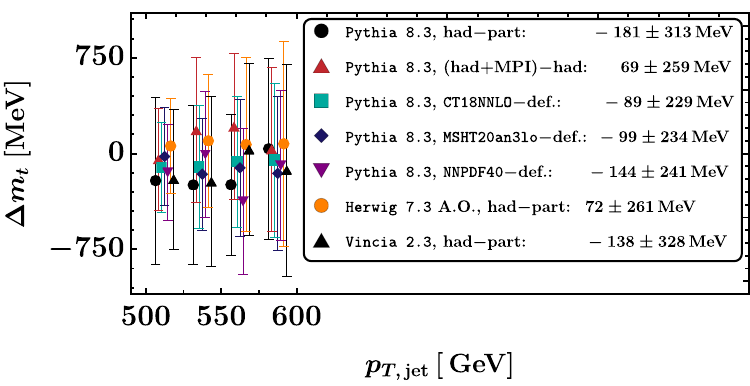}
\label{fig:conversion_b}}
\caption{The errors shown in panels (a) and (b) are conservatively computed from the cumulant of statistical errors, reasonable variation in the polynomial degree used for the peak fit, and variation of the peak fit range by $\pm$ 10$\%$. Jet $p_{T}$ is in 25 GeV bins. Horizontal clusters of points all share the same central value of $p_{T, \, \mathrm{jet}}$.}
\label{fig:conversion}
\end{figure}

The conversion should be performed using a full calculation of the observable. The formalism to achieve this will be presented in future work. Here, we show how this can be performed using a parton shower study where a polynomial is used to fit for the peak positions $\zeta_t$ and $\zeta_W$. In the large boost limit of a top quark decay~\cite{Holguin:2024tkz},
\begin{align}\label{eq:Cfactor}
m_{t}= m_{W} \left[ C(\as,R) \sqrt{\zeta_{t}/\zeta_W} + \mathcal{O}\left(\frac{m_{W}}{ p_{T,{\rm jet}} }, \frac{m_{t}}{ p_{T,{\rm jet}}} \right)\right] ,
\end{align}
where the coefficient $C$ has a perturbative expansion and depends on the jet radius $R$. $C$'s magnitude is dominated by the relative boost of the $W$ boson which is largely determined by the top decay at fixed order. The exact magnitude of $C$ will differ between event generators which employ different approximations to the NLO top decay. We considered the \Herwigxx~\cite{Bahr:2008pv,Bahr:2008tf,Bellm:2019zci}, \Pythiaxx~\cite{Sjostrand:2014zea} and \Vinciaxx~\cite{vincia:2016} parton showers. We reconstruct anti-$k_T$~\cite{Cacciari:2008gp} jets with radii $R=1, ~ 1.2, ~ 1.5$ using \Fastjet~\cite{Cacciari:2011ma}. We found the extraction of the top mass to be largely independent of $R$, and so we here focus on $R=1.2$. We first illustrate that the ratio $\zeta_t/\zeta_W$ extracted from the peak positions is virtually independent of the jet $p_T$. We then show that this ratio is largely insensitive to nonperturbative effects.

In \fig{conversion_a}, we show the ratio of the $W$ and top peaks, $\zeta_t/\zeta_W$ at parton level, hadron level, and with MPI (\Pythia model for UE) turned on, as well as for a variety of different PDFs. While perturbative corrections can be incorporated to further improve the conversion, \fig{conversion_a} shows that we have eliminated the dominant $p_T$ scaling from the problem by using the standard candle approach. We emphasize that the location of the top and $W$ peaks are themselves strongly dependent on the choice of the PDF set, which influences the jet $p_T$ distribution. Still, this dependence cancels in the ratio to a remarkable degree. We note that \Herwig confirms the cancellation of the $p_{T}$ dependence but differs by $3 \%$ from every other Monte Carlo in the perturbative constant dictating the scale of the $y$-axis, specifically $C$ in \eq{Cfactor}.
\adi{As elaborated in the \textit{Supplemental Material},}
the difference is expected to originate from the absence of NLO top decays in \Pythiaxx and \Herwigxx \cite{Holguin:2024tkz} and is readily improvable \cite{Campbell:2012uf,Chen:2022wit,Chen:2023dsi,Jezo:2023rht}. In \fig{conversion_b}, we show the size of the shift in the extracted value of the top mass due to hadronization, MPI, and variation of the PDF set. All are observed to be $\lesssim 200$ MeV, and almost all are consistent within errors with no shift (see also \cite{Holguin:2024tkz}). This is very promising compared with shifts of the order of $\sim 1$ GeV due to MPI in the groomed jet mass \cite{Hoang:2017kmk}, and illustrates the robustness of our observable. We stress that the magnitude of these shifts should not be taken as a source of uncertainty due to nonperturbative effects since we expect that the leading nonperturbative corrections can be understood field theoretically \cite{Belitsky:2001ij,Korchemsky:1999kt,Korchemsky:1997sy,Schindler:2023cww}. It should instead be taken as a demonstration of insensitivity to physics at and below the shower cutoff, illustrating that the observable is not sensitive to physics below $\Gamma_t$ as desired.

\emph{Feasibility at the HL-LHC.}---Extracting the top mass from our proposed EEC-based observable will require the measurement of the distribution discussed above and an understanding of the associated experimental uncertainties, as well as a theoretical calculation expressing the distribution in terms of the parameter $m_t$.
The goal of this \emph{Letter} is
to introduce an observable for precision top mass extractions and illustrate that it has the desired properties so as to motivate a dedicated experimental and theoretical effort. To do so, we will illustrate the statistical feasibility of our approach at the LHC and perform a parton-shower-based extraction of the top mass using polynomial fits of the peak positions.
More details of the experimental aspects of this study are presented in a longer companion paper \cite{Holguin:2024tkz}. We performed an estimate of the statistical uncertainty at Run 2 and 3 of the LHC and at the HL-LHC using pseudodata from \Pythiaxx. For the Run 2 pseudodata set, we generated jets
corresponding to the number of selected jets in the CMS Run 2 top jet mass measurement~\cite{CMS:2022kqg} (52000 top jets). For the
Run 3 and HL-LHC pseudodata sets the number of jets is increased by factors of 300/138 and 3000/138,
assuming integrated luminosities of 300fb$^{-1}$ and 3000fb$^{-1}$ respectively. Our results forecast statistical precision on the top mass better than $\sim 300$ MeV for the HL-LHC, which is extremely promising. We also find that this measurement is statistically feasible with present LHC luminosities with about 1 GeV or better precision.

The observable we propose can also be computed on charged particles only (tracks) \cite{Chang:2013rca,Chang:2013iba,Jaarsma:2023ell,Chen:2022pdu,Chen:2022muj,Jaarsma:2022kdd,Li:2021zcf}, which provide considerably better angular resolution. It will be interesting to study the interplay of the increased angular resolution against the reduced statistics by incorporating estimates of the tracking efficiency.

\emph{Outlook.}---We have focused on introducing our EEC-based observable and illustrating its desirable properties through parton-shower analyses. We believe that our results provide a strong motivation for its systematic theoretical study. Here, we emphasize that resummation in the peak region, which is required for precise theory predictions, is achievable with high perturbative accuracy.
As the observable in question is more complex than the jet mass, due to the fact that it involves the measurement of angular correlations on the radiation in the top decay at the scale $\Gamma_t$, the development of rigorous factorization theorems for EECs on unstable states is required. This can be achieved by combining HQET~\cite{Isgur:1989vq,Georgi:1990um,Korchemsky:1991zp}, SCET~\cite{Bauer:2001ct,Bauer:2000yr,Bauer:2001yt}, and on-shell particle EFT~\cite{Beneke:2003xh,Beneke:2004km,Beneke:2015vfa} with the factorization theorems for the EECs in \Refscite{Moult:2018jzp,Gao:2019ojf}. These give rise to a variety of interesting field theoretic structures, which will be explored in future work. Still, it is important to stress that many of the ingredients are already known to high perturbative orders.
A key theoretical feature is that, in the peak region, the distributions we are interested in are described by SCET$_{\rm{II}}$-type factorization theorems, with the lowest scale being the top width $\Gamma_t$. This is in contrast to the jet mass, which is described by a SCET$_{\rm{I}}$ framework, where the scale of soft radiation is parametrically lower than $\Gamma_t$. This also emphasizes the complementarity of the jet mass and EEC approaches.

\emph{Conclusions.}---In this \emph{Letter}, we have introduced a top-mass-sensitive energy correlator (EEC) observable with desirable experimental and theoretical properties. The key to our proposal is to simultaneously extract the angular scales associated with $m_W$ and $m_t$ and to use the $W$ mass as a standard candle to calibrate the top mass. This allows us to overcome previous issues of EEC-based approaches, which require the knowledge of the jet $p_T$ to convert between angular and mass scales~\cite{Holguin:2022epo}. We performed a preliminary study using luminosity estimates for the LHC, finding promising results.

The discovery of a top-mass-sensitive observable with robust experimental and theoretical properties in a hadron collider environment is a crucial addition to the physics program of the LHC.
Our proposed observable opens up a rich theoretical and experimental program, enabling a variety of new theoretical techniques to be applied to the phenomenology of heavy particle decays.

\emph{Acknowledgements.}---We thank Gregory Korchemsky and Iain Stewart for useful discussions. I.M. is supported by start-up funds from Yale University. J.H. is supported by the Leverhulme Trust as an Early Career Fellow. This work is supported in part by the GLUODYNAMICS project funded by the ``P2IO LabEx (ANR-10-LABX-0038)'' in the framework ``Investissements d'Avenir'' (ANR-11-IDEX-0003-01) managed by the Agence Nationale de la Recherche (ANR), France.
A.P. acknowledges support from DESY (Hamburg, Germany), a member of the Helmholtz Association HGF.
We are grateful to Simon Pl\"atzer for technical support with \Herwig. We thank Patrick Komiske who developed the \texttt{EnergyEnergyCorrelators} package used in our EEC simulations \cite{EEC_github}.

\bibliography{EEC_ref.bib}{}
\bibliographystyle{apsrev4-1}

\newpage
\onecolumngrid
\newpage
\newpage

\section{Supplemental material}

\subsection{Comparing results from Herwig 7.3 and 7.2 versions}

\begin{figure}[h!]
\centering
\includegraphics[width=0.47\textwidth]{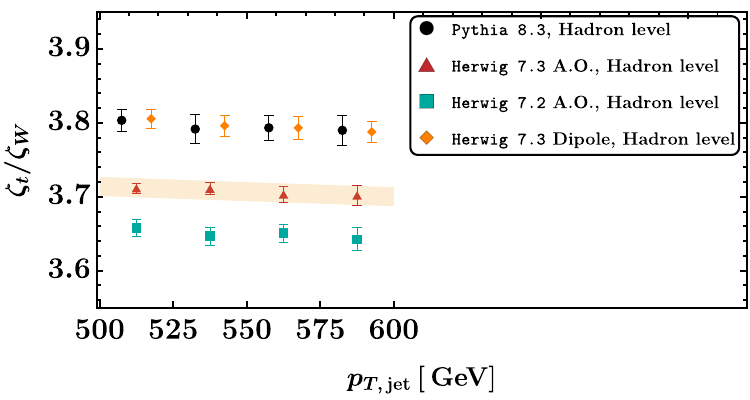}
\caption{The same as \fig{conversion_a} but with the comparison between \Herwigxx and \Herwig\texttt{7.2}. The orange band around \Herwigxx angular ordered shower data points shows the size of $\Delta m_t = \pm 300$ MeV variation in the top mass due to effects such as hadronization, FSR scale, PDFs and the underlying event. As discussed below, the significantly larger difference between \Herwigxx angular ordered shower and dipole showers results from differences in the perturbative approximations of the NLO corrections to the top quark decay.}
\label{fig:suppA}
\end{figure}

\begin{figure}[h!]
\centering
\includegraphics[width=0.47\textwidth]{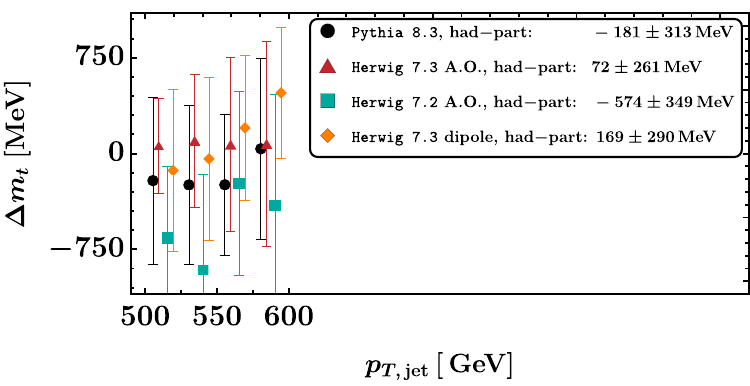}
\caption{ The same as \fig{conversion_b} but with the comparison between \Herwigxx and \Herwig\texttt{7.2}.}
\label{fig:suppB}
\end{figure}

We now compare the results using \Herwig\texttt{7.2} (which were presented in version 1 of this manuscript) with those using \Herwigxx.
Three different \Herwig models are compared against \Pythia in Figures \ref{fig:suppA} and \ref{fig:suppB}. Each of the \Herwig models displays the key features that we desired to see in our analysis: specifically that the $p_T$ dependence of $\zeta_t / \zeta_W$ has largely canceled and that corrections from hadronization are small.
However, \Herwig\texttt{7.2} also displays some other notable differences from \Pythia.
The hadronization corrections are somewhat erratic and noticeably larger in the $p_{T,{\rm jet}}\sim 500-550~$GeV range than in the $550-600~$GeV range. Additionally, the scale factor of the $y$-axis in \fig{suppA} differs between \Herwig\texttt{7.2} and \Pythia. However, a comparison with \Herwigxx angular ordered and dipole showers shows that these effects are spurious in nature. In both \Herwigxx models, the hadronization corrections (\fig{suppB}) are significantly reduced with respect to \Herwig\texttt{7.2} and are no longer erratic, making them as small as in \Pythia. With regard to the difference in the scale of the $y$-axis of \fig{suppA}, the \Herwigxx dipole shower is in complete agreement with the \Pythia shower and the \Herwigxx angular ordered shower has moved closer to the dipole shower results. Nevertheless, a small discrepancy in the y-axis scale still persists between the \Herwigxx angular ordered shower the dipole showers, which we will discuss in the following section.

\subsection{Dissecting the discrepancy between Herwig angular ordered shower and other parton showers}

Here, we show that the apparent discrepancy observed between $\zeta_t/\zeta_W$ ratio predictions using \Herwigxx angular ordered shower and other showers in \fig{conversion_a} of the \letter and \fig{suppA} of the \emph{supplemental material} is of purely perturbative origin.
Let us summarize the dipole/antenna shower results presented in \fig{conversion_a} and \fig{suppA} through the following equation:
\begin{align}\label{eq:dipolePart}
\big(\zeta_t/\zeta_W\big)^{\rm had.}_\text{dipole \hbox{-}shower} \sim
\big(\zeta_t/\zeta_W\big)^{\rm had. \, (cluster)}_\text{\Herwigxx dipole} \sim \big(\zeta_t/\zeta_W\big)^{\rm had. \,(string)}_\text{\Pythia} \sim \big(\zeta_t/\zeta_W\big)^{\rm had.\, (string)}_\text{\Vincia} \, ,
\end{align}
where the $\sim$ here refers to equality of the results up to few hundred MeV differences in the resulting top mass. The superscript `had.' refers to shower predictions at hadron-level, and we additionally indicate the specific hadronization model in brackets: i.e. cluster fission in \Herwig and string breaking in \Pythia/\Vincia.
Now, from \fig{conversion_a} or equivalently \fig{suppA}, the \Herwigxx angular ordered result differs from the dipole shower results by $\sim$ 2\%:
\begin{align}\label{eq:AOPart}
\Big | \big(\zeta_t/\zeta_W\big)^{\rm had.}_\text{\Herwigxx A.O.} - \big(\zeta_t/\zeta_W\big)^{\rm had.}_\text{dipole \hbox{-}shower} \Big | \sim 0.02 \times \big(\zeta_t/\zeta_W\big)^{\rm had.}_\text{dipole \hbox{-}shower} \, .
\end{align}
While this discrepancy is notable, it can be completely understood as having a purely perturbative origin which can be brought under theoretical control. The discrepancy cannot be ascribed to hadronization modelling, and thus poses no problem for our proposed measurement. This statement is justified by our analysis as follows. Let us summarize the results of \fig{conversion_b} of the \letter:
\begin{align}\label{eq:allHad}
&\Big| \big(\zeta_t/\zeta_W\big)^\text{had. (cluster)}_\text{\Herwigxx dipole} -\big(\zeta_t/\zeta_W\big)^{\rm part.}_\text{\Herwigxx dipole} \Big | & &\longrightarrow& &\Delta m_t = + 169 ~ \pm ~ 290 \,\text{MeV}&
\\ \nn
&\Big| \big(\zeta_t/\zeta_W\big)^\text{had. (cluster)}_\text{\Herwigxx A.O.} -\big(\zeta_t/\zeta_W\big)^{\rm part.}_\text{\Herwigxx A.O.} \Big | & &\longrightarrow& &\Delta m_t = +72 ~ \pm ~ 261 \,\text{MeV}&\\
\nn &\Big| \big(\zeta_t/\zeta_W\big)^\text{had. (string)}_\text{\Pythiaxx} -\big(\zeta_t/\zeta_W\big)^{\rm part.}_\text{\Pythiaxx} \Big | & &\longrightarrow& &\Delta m_t = -181 ~ \pm ~ 313 \,\text{MeV}&\\
&\Big| \big(\zeta_t/\zeta_W\big)^\text{had. (string)}_\text{\Vinciaxx} -\big(\zeta_t/\zeta_W\big)^{\rm part.}_\text{\Vinciaxx} \Big | & &\longrightarrow& &\Delta m_t = -138 ~ \pm ~ 328 \,\text{MeV}& \nn \, .
\end{align}
With \eqs{dipolePart}{allHad}, we can safely conclude that
\begin{align}
\Big| \big(\zeta_t/\zeta_W\big)^\text{had.} -\big(\zeta_t/\zeta_W\big)^{\rm part.} \Big | ~~~~ \longrightarrow ~~~~ \Delta m_t\lesssim 300\, \text{MeV}\,.
\end{align}
Thus, the results above imply that the proposed top mass observable is resilient against hadronization effects whichever way these are modeled. This is crucially true for the \Herwigxx angular ordered prediction, demonstrating that the deviation in \eq{AOPart} is not compensated for by hadronization modelling. For better clarity, the orange band in \fig{suppA} around the \Herwigxx A.O. points illustrates the size of $\Delta m_t =\pm 300$ MeV top mass variations in the $\zeta_t/\zeta_W$ ratio, which is significantly smaller than the difference with dipole showers.

In \figs{conversion_a}{conversion_b} we similarly break down other components between each of the generators, for instance the MPI modeling and the PDF sets. Additionally, in \Refcite{Holguin:2024tkz}, we take this yet further and look at parameter tunes, scale variations, hard process NLO matching, $b$-fragmentation modeling, parton shower recoil schemes, color reconnection, the modeled top decay phase-space, and further experimental considerations. In every instance, we find that effects are less than $300\, \text{MeV}$ --- indeed they are mostly of the order $100 \,$MeV. Particularly noteworthy is that even the shift due to FSR scale variations are of order $200 \,$MeV, which approximates the expected error due to the resummations performed by the parton showers. Therefore our results systematically rule out any explanation for the difference in \eq{AOPart} other than the perturbative description of the top decay. This decay is only handled at LO in all the event generators considered, with the parton shower approximating the NLO decay. We therefore can deduce that an angular ordered shower provides a different prediction only because it differently populates the NLO fixed order phase-space.\footnote{ This is common cause of discrepancies between angular-ordered showering vs $k_t$ (dipole/antenna) showers. For instance, the angular ordered shower under-populates the 3-jet tail of the the thrust distribution ($T \lesssim 0.66$, see for example \cite{Bewick:2019rbu}) without NLO fixed order matching.} We stress that our analysis demonstrates that our proposed method to extract the top quark mass is not crucially affected by hadronization, underlying event, soft/Coulomb gluon resummation, color reconnection, PDF uncertainties etc. that have been bottlenecks in other approaches.

\subsection{On subleading $\mb{p_T}$ dependence of $\mb{\zeta_t/\zeta_W}$}

In \eq{Cfactor}, the extraction of the top mass from $\zeta_t/\zeta_W$ is presented at leading power in the large-boost limit, and it is found to be independent of the jet $p_T$. However, at sub-leading power, $\zeta_t/\zeta_W$ does depend on the jet $p_T$. We can summarize the implications for our proposed measurement as follows: while $\zeta_t \sim (m_t/p_T)^2$—which, on its own, could introduce an uncertainty on the extracted top mass of the order of a few GeV—the ratio $\zeta_t/\zeta_W \sim (m_t/m_W)^2 + \mathcal{O}((m_W/p_T)^3)$ exhibits a very strongly suppressed dependence on the jet $p_T$, as this dependence is eliminated at leading power. This reduces the uncertainty from determining the jet $p_T$ from a few GeV to approximately $100 \,$MeV. In the following discussion, we will provide quantitative estimates for the size of the sub-leading corrections.

The small residual $p_T$ dependence in the ratio $\zeta_t/\zeta_W$ is not limiting for our proposed measurement for two reasons:
\begin{enumerate}[label=(\alph*)]
\item There is a systematic procedure to incorporate power corrections in perturbative calculations which describe this $p_T$ dependence. So, we can expect to describe the slope in $p_T$ robustly in perturbation theory.
\item With the leading $p_T$ dependence canceled, any undesirable shifts in $p_T$ due to PDFs, underlying event contribution, color reconnection, hadronization, etc. are highly suppressed to a few hundred MeV level, as demonstrated in a number of simulation studies we performed in this \letter and in \Refcite{Holguin:2024tkz}.
\end{enumerate}

\begin{figure}[t]
\centering
\includegraphics[width=0.47\textwidth]{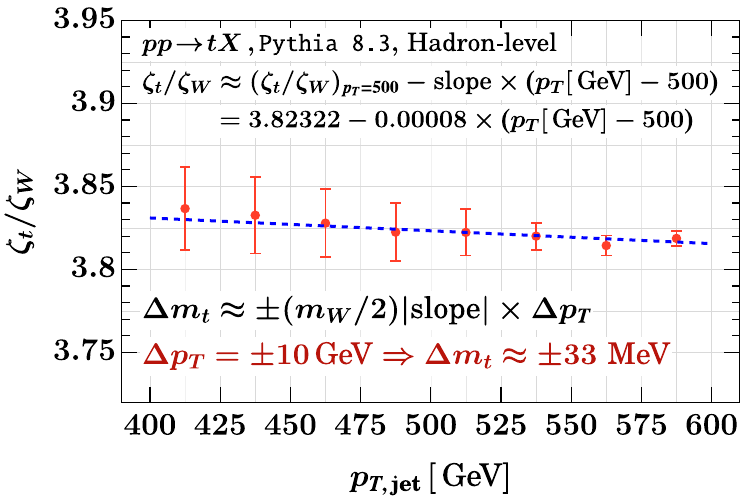}
\includegraphics[width=0.47\textwidth]{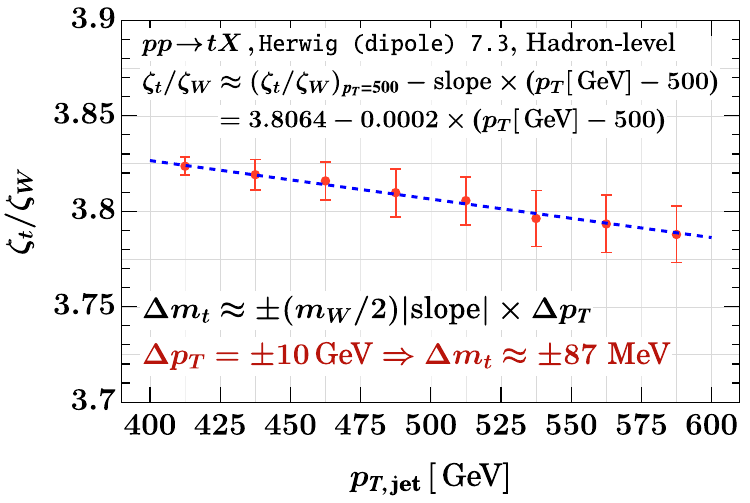}
\caption{The ratio $\zeta_t/\zeta_W$ shown for wider range of $p_{T,\rm jet}$ for hadron level \Pythiaxx and \Herwigxx dipole shower simulations for $R=1.2$ radius jets. The ratio $\zeta_t/\zeta_W$ exhibits a small slope in $p_T$ due to subleading power correction. However, the impact of uncertainties associated with $p_T$ through this residual dependence in the extracted top mass is negligible.
Here $C^{(1.2)}_{\rm MC} \sim 1$, see \cite{Holguin:2024tkz} for further details.}
\label{fig:suppC}
\end{figure}

Let us now justify (b) quantitatively. In the large boost limit, the peak locations $\zeta_t$ and $\zeta_W$ can in general be written as power series in $m_t/p_T$ and $m_W/p_T$. Schematically, by dimensional analysis,
\begin{align}\label{eq:zetatw}
\zeta_t &= \bigg(\frac{m_t^2}{p_T^2}\bigg) \Bigg( a_0+ a_1 \frac{m_t}{p_T} +a_2 \bigg(\frac{m_t}{p_T}\bigg)^2 + \ldots\Bigg) \,, \\
\zeta_W &= \bigg(\frac{m_W^2}{p_T^2}\bigg) \Bigg( b_0+ b_1 \frac{m_W}{p_T} +b_2 \bigg(\frac{m_W}{p_T}\bigg)^2 + \ldots\Bigg) \nn \,,
\end{align}
where the the coefficients $a_i$ and $b_i$ are $\cO(1)$ numbers independent of $p_T$, and the higher-order terms are power suppressed. In previous work~\cite{Holguin:2022epo}, it was proposed that the top mass could be directly extracted from $\zeta_t$. However, effects such as hadronization, UE, uncertainties in PDFs, limitations of unfolding the $p_T$ distribution, etc., can induce uncertainties in the knowledge of the jet $p_T$ by $\Delta p_T \sim 10$ GeV (see, for example, Fig.~(11) of \Refcite{Holguin:2022epo}). It was acknowledged that, if left unaccounted for, this could leave a $2$-$3\,$GeV uncertainty on the extracted top mass. Now let us consider the ratio of $\zeta_t/\zeta_W$:
\begin{align}
\frac{\zeta_t}{\zeta_W} = \frac{a_0}{b_0} \frac{m_t^2}{m_W^2} \Bigg(1+ \left(\frac{a_1 m_t}{a_0 m_W} - \frac{b_1}{b_0}\right) \frac{m_W}{p_T} + \cO \bigg(\frac{m_W^2}{p_T^2}\bigg) \Bigg)\,,
\end{align}
which can be re-arranged into a top mass extraction as:
\begin{align}
m_t = m_W\, C \,\sqrt{\frac{\zeta_t}{\zeta_W}} \Bigg(1 + D \frac{m_W}{p_T} + \cO\bigg(\frac{m_W^2}{p_T^2}\bigg) \Bigg)\,.
\end{align}
Here $C= \sqrt{b_0/a_0} \approx 1$ is the coefficient $C$ used in \eq{Cfactor} and $D$ is an $\cO(1)$ number that is a function of $m_t/m_W$, $a_{0,1}$ and $b_{0,1}$.
We can extract $D$ from the residual jet $p_T$ dependence in $\zeta_t/\zeta_W$, such as is seen in Fig. \ref{fig:conversion_a}. Additionally, a wider ranger of $p_{T,\rm jet}$ was extensively studied in \cite{Holguin:2024tkz}. The combination $m_W C\sqrt{\zeta_t/\zeta_W}$ varies roughly by $1 \pm 0.5$~GeV when $p_T$ is varied between $[400,600]$ GeV~\cite{Holguin:2024tkz}. This results in $D = 0.1 \pm 0.05$.

We can now consider propagating the $\Delta p_{T} \sim 10\,$GeV uncertainty on a top jet with $p_T=500\,$GeV to our proposed measurement. We find that,
\begin{align}
\Delta m_t = C\,\sqrt{\frac{\zeta_t}{\zeta_W}} \,D \, \frac{m^2_W}{p^2_T} \Delta p_{T} \approx m_t \, D\, \frac{m_W}{p^2_T} \Delta p_{T} \lesssim 100 \mathrm{MeV}\,.
\end{align}
In \fig{suppC} we show that the the impact of uncertainties in $p_T$ on the extracted top mass for \Pythiaxx and \Herwigxx hadron-level simulations. In each case, we observe a significant suppression of $p_T$-induced uncertainties which is compatible with the expected $\cO(200\,{\rm MeV})$ level uncertainties seen in \Refcite{Holguin:2024tkz}.

Furthermore, our measurement can be performed across multiple $p_T$ bins, thereby constraining uncertainties in the measurement yet further.
This size of intrinsic uncertainty is indeed what we have observed in the exhaustive variations we presented across this \letter (and the supporting paper, \Refcite{Holguin:2024tkz}). We can conclude that our proposed measurement is indeed very successful at removing the intrinsic uncertainties present in the jet $p_T$ distribution from the extraction of the top mass.

\end{document}